\documentclass[a4paper,
               keeplastbox,   % flushend option: not to un-indent last line in References
               ]{jacow}
\usepackage{pdfpages,multirow,ragged2e} %
\usepackage{amsmath} 

\makeatletter
\newcommand{\mathleft}{\@fleqntrue\@mathmargin0pt}
\newcommand{\mathcenter}{\@fleqnfalse}
\makeatother

\makeatletter%
	\ifboolexpr{bool{xetex}}
	 {\renewcommand{\Gin@extensions}{.pdf,%
	                    .png,.jpg,.bmp,.pict,.tif,.psd,.mac,.sga,.tga,.gif,%
	                    .eps,.ps,%
	                    }}{}
\makeatother

\ifboolexpr{bool{xetex} or bool{luatex}} % test for XeTeX/LuaTeX
 {}                                      % input encoding is utf8 by default
 {\usepackage[utf8]{inputenc}}           % switch to utf8

\usepackage[USenglish]{babel}
\usepackage{multicol,lipsum}
\usepackage{cuted}
\ifboolexpr{bool{jacowbiblatex}}%
 {%
  \addbibresource{jacow-test.bib}
  \addbibresource{biblatex-examples.bib}
 }{}
\begin{document}

\title{Beam-beam blowup in the presence of x-y coupling sources at FCC-ee}

\author{D. El Khechen\thanks{dima.el.khechen@cern.ch}, K. Oide\textsuperscript{1}, F. Zimmermann, CERN, Geneva, Switzerland \\
		\textsuperscript{1} also at High Energy Accelerator Research Organisation, Tsukuba, Ibaraki, Japan }
	
\maketitle

\begin{abstract}
  FCC-ee, the lepton version of the Future Circular Collider (FCC), is a 100 Km future machine under study to be built at CERN. It acquires two experiments with a highest beam energy of 182.5 GeV. FCC-ee aims to operate at four different energies, with different luminosities to fulfill physics requirements. Beam-beam effects at such a high energy/luminosity machine are very challenging and require a deep understanding, especially in the presence of x-y coupling sources. Beam-beam effects include the beamstrahlung process, which limits the beam lifetime at high energies, as well as dynamic effects at the Interaction point (IP) which include changes in the beta functions and emittances. In this report, we will define the beam-beam effects and their behaviours in the FCC-ee highest energy lattice after introducing x-y coupling in the ring.
  \end{abstract}

\section{Introduction}
Beam-beam effects including beamstrahlung have been studied for FCC-ee. Dynamic effects including dynamic $\beta$ functions and dynamic emittances were simulated in the presence of vertical misalignments of the sextupoles. Tracking was performed in SAD \cite{sad} with a beam-beam element present at both IPs of the FCC-ee highest energy lattices (175 and 182.5 GeV). The beam-beam is represented by a weak-strong beam beam simulation (BBWS) \cite{bbws} which is implemented in SAD.
Beam blowup was observed by tracking in the presence of beam-beam and without beam beam. We will report on the different simulations and discuss the results.
\section{Dynamic $\beta$ functions and emittances}
Dynamic effects are the change of the Twiss parameters ($\beta$ functions and emittances) at the IP due beam beam quadrupolar focusing. These dynamic effects are enhanced by running at half integer or integer resonance tunes and thus affecting the luminosity.
\subsection{Analytical estimations}
Dynamic beta functions can be easily calculated by the half turn matrix as given in Eq.\ref{dynamic beta}, where $\beta$, $\beta_0$, $\mu$, $\mu_0$ are the dynamic $\beta$, the design $\beta$, the shifted betatron tune after beam beam and the design betatron tunes respectively and $\frac{1}{f}$ being the beam beam strength. Solving Eq. \ref{dynamic beta}, we can obtain the dynamic beta function given in Eq.\ref{beta} where $\xi_{x,y}$ is the so-called beam beam parameter and expressed in terms of $\beta$ function and beam-beam force as given in Eq.\ref{beam beam}

%\begin{widetext}

\begin{equation}
\begin{split}
\begin{pmatrix} 
 \cos\mu & \beta \sin\mu \\
-\frac{1}{\beta}\sin\mu & \cos\mu 
\end{pmatrix} 
=
\begin{pmatrix} 
1 & 1 \\
-\frac{1}{2f}& 0
\end{pmatrix} 
\begin{pmatrix} 
\cos\mu_0 & \beta_0 \sin\mu_0 \\
-\frac{1}{\beta_0}\sin\mu_0 & \cos\mu_0 
\end{pmatrix} 
\begin{pmatrix} 
1 & 1 \\
-\frac{1}{2f}& 0
\end{pmatrix} 
\label{dynamic beta}
\end{split}
\end{equation}

%\end{widetext}

\begin{equation}
\beta=\frac{\beta_{x,y}}{\sqrt{1-(2\pi\xi_{x,y})^2+4\pi\xi_{x,y}cot(\mu_0{_{x,y}})}}
\label{beta}
\end{equation}

\begin{equation}
\xi_{x,y}=\frac{\beta_0{_{x,y}}}{4\pi f_{x,y}}
\label{beam beam}
\end{equation}

Analytical estimations of dynamic emittance require longer calculations. One way to calculate the dynamic horizontal emittance is given in \cite{dynamic emittance}. The calculation of the dynamic vertical emittance is not straight forward since it requires the knowledge of the errors and corrections of the lattice.

\subsection{Simulations of dynamic effects}

The dynamic $\beta$ functions and emittance were also simulated by an insertion of a thin quadrupole at both IPs representing the linear beam-beam. The thin quadrupole gives a kick as given in Eq.\ref{kick}. 
\begin{equation}
KL=\frac{4\pi f_{x,y}}{\beta_{0_{x,y}}}
\label{kick}
\end{equation}

The beam beam parameters in this case are given to be (0.095,0.157) in the horizontal and vertical plane respectively. It is important to mention that the vertical emittance in the lattice is generated by vertically misaligned sextupoles to achieve an xy coupling of 0.2\%. Furthermore, the values of the betatron tunes for the given lattice are ($\nu_x$,$\nu_y$)=(0.553, 0.59). The new values of $\beta$ functions and emittance, after the insertion of the thin quadrupole, are then extracted.
The results of the simulations for the dynamic $\beta$ perfectly match with the analytical estimations and the results of the dynamic effects are summarized in the Table.\ref{dynamic table}. It is important here to highlight that the values of the dynamic emittance depend essentially on the way the xy coupling is introduced in the lattice.

\begin{table}[!hbt]
   \centering

   \begin{tabular}{lcc}
       \toprule
       \textbf{Twiss functions} & \textbf{Design values}                      & \textbf{Dynamic values} \\
       \midrule
           $\beta_x$ (m)        & 1         & 0.49        \\ %[3pt]
           $\beta_y$ (mm)       & 2           & 1.12    \\ %[3pt]
            $\epsilon_x$ (nm)       & 1.34          & 2.14       \\ %[3pt]
             $\epsilon_y$  (pm)    & 2.7          & 3.83       \\
       \bottomrule
   \end{tabular}
      \caption{Values of dynamic $\beta$ functions and emittances for one given random generator}
   \label{dynamic table}
\end{table}

\section{Tracking and blowup}
\subsection {Tracking with beam beam}
\begin{figure}[!tbh]
 \includegraphics[width=\columnwidth]{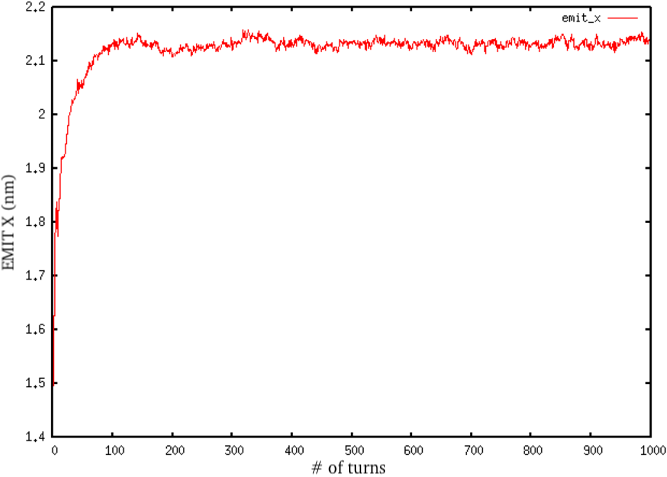}
 \caption{Evolution of horizontal emittance in the presence of the beam beam element}
  \label{fig:emitx}
    \end{figure}

\begin{figure}[!tbh]
        \includegraphics[width=\columnwidth]{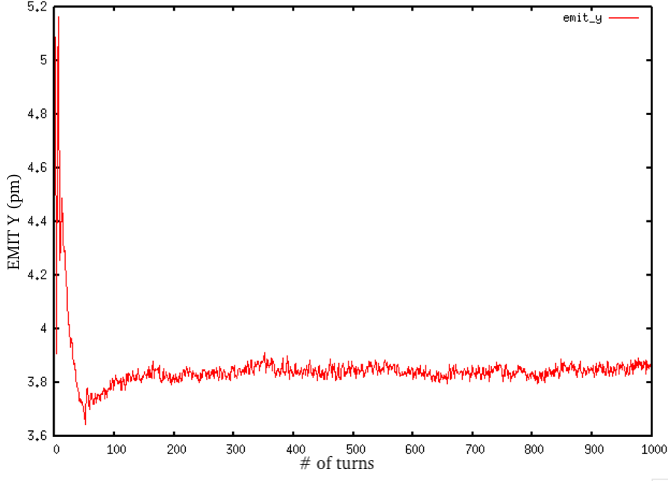}
         \caption{Evolution of vertical emittance in the presence of the beam beam element}
  \label{fig:emity}
    \end{figure}

Tracking was also considered to study the beam beam dynamic effects. This was performed by inserting two beam beam elements at both IPs of the lattice. Tracking of 10$^{4}$ macroparticles  beam initially generated at the IP was performed over 500 turns (50 turns= 1 longitudinal damping period). Every turn, the $\beta$ functions and the emittances of the tracked beam were calculated at the IP.  In the following section, the vertical emittance is introduced by vertically misaligning the sextupoles to achieve an xy coupling of 0.2\%. For the time being, only one seed is used for the random misalignments of the sextupoles.
Tracking in the presence of the beam beam elements showed a blowup in the vertical (from 2.7 pm to 3.83 pm) and horizontal (from 1.34 nm to 2.14 nm) emittances as shown in Figs.(\ref{fig:emitx},\ref{fig:emity})
Such a blowup in the vertical emittance was not observed in the simulations without misalignments\cite{Demin}. To verify the source of the blowup, a tracking was performed without a beam beam element.

\subsection{Tracking without beam beam}
\begin{figure}[!tbh]
        \includegraphics[width=\columnwidth]{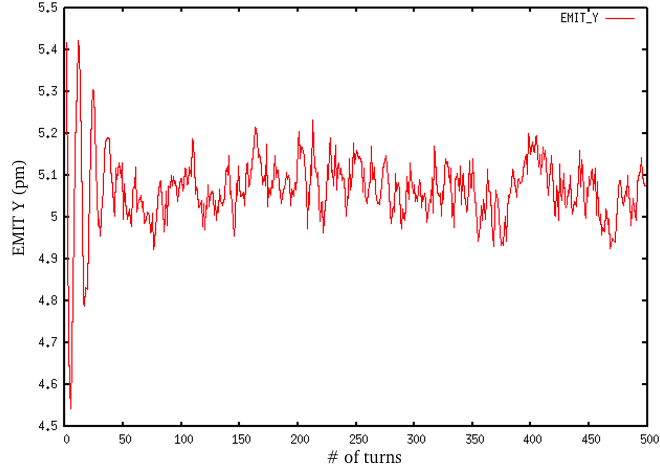}
         \caption{Evolution of the vertical emittance in the absence of beam beam element}
  \label{fig:emity2}
    \end{figure}

Beam beam elements are removed and a simple tracking was performed in the presence of vertical misaligned sextupoles to achieve 2.7 pm vertical emittance. The same seed used for the sextupole misalignments for tracking with the beam beam was also used for this tracking. Unexpected emittance blowup of the vertical emittance was observed (from 2.7 pm to 5 pm). This blowup is shown in Fig. \ref{fig:emity2}. A possible reason for such a blowup could be related to the dispersions and xy coupling induced by the sextupole misalignments in the ring. 

\section{More samples tracking}
For a better understanding, different seed generators for sextupole misalignments have been produced and tracking was performed with and without the beam beam element. Results showed the clear dependence of the blowup on the seed number.
\subsection{With beam beam}
\begin{figure}[!tbh]
        \includegraphics[width=\columnwidth]{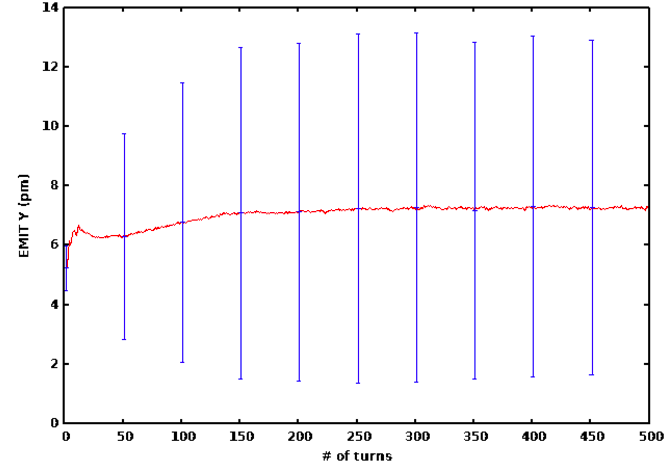}
         \caption{The evolution of the average vertical emittance over twelve samples in the presence of beam beam element  }
  \label{fig:averagebeambeam}
    \end{figure}

\begin{figure}[!tbh]
        \includegraphics[width=\columnwidth]{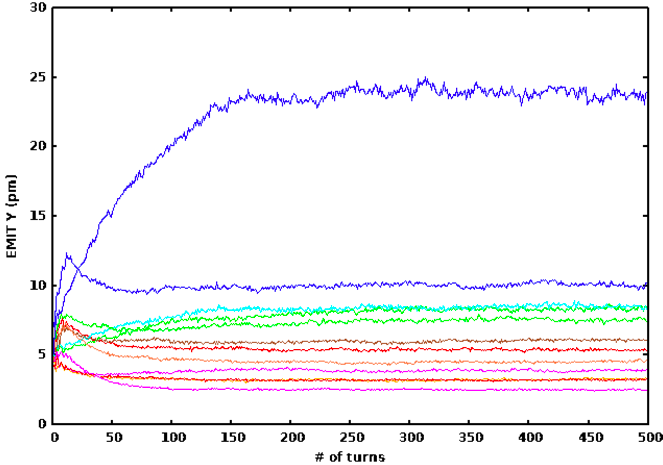}
         \caption{The evolution of the vertical emittance of the twelve different samples in the presence of the beam beam element  }
  \label{fig:12beambeam}
    \end{figure}

Twelve different samples were simulated and tracking of the beam was performed in the presence of the beam beam element, over 500 full turns. The vertical emittance was then averaged over the different samples and the result is shown in Fig. \ref{fig:averagebeambeam} along with error bars. The error bars are big, this is explained by the variation of the amount of the vertical emittance blowup for different random generator seeds shown in Fig. \ref{fig:12beambeam}.

\subsection{Without beam beam}
\begin{figure}[!tbh]
        \includegraphics[width=\columnwidth]{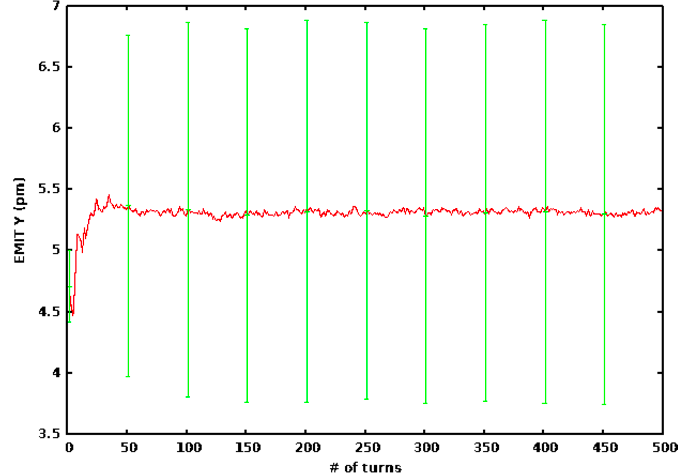}
         \caption{The evolution of the average vertical emittance over twelve samples in the absence of beam beam element  }
  \label{fig:averagenobeambeam}
    \end{figure}

\begin{figure}[!tbh]
        \includegraphics[width=\columnwidth]{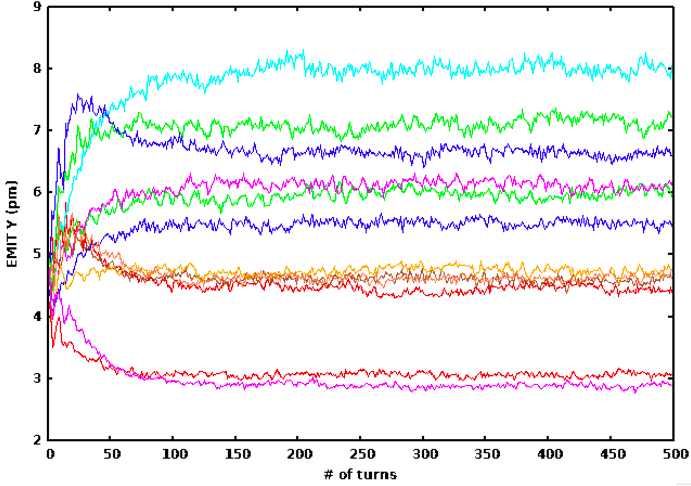}
         \caption{The evolution of the vertical emittance of the twelve different samples in the absence of the beam beam element  }
  \label{fig:12nobeambeam}
    \end{figure}

The same procedure was done in the absence of the beam beam element, the result of the average emittance is shown in Fig. \ref{fig:averagenobeambeam}. The individual seeds are shown in Fig. \ref{fig:12nobeambeam}.

\subsection{Compare Optics}

\begin{figure}[!tbh]
        \includegraphics[width=\columnwidth]{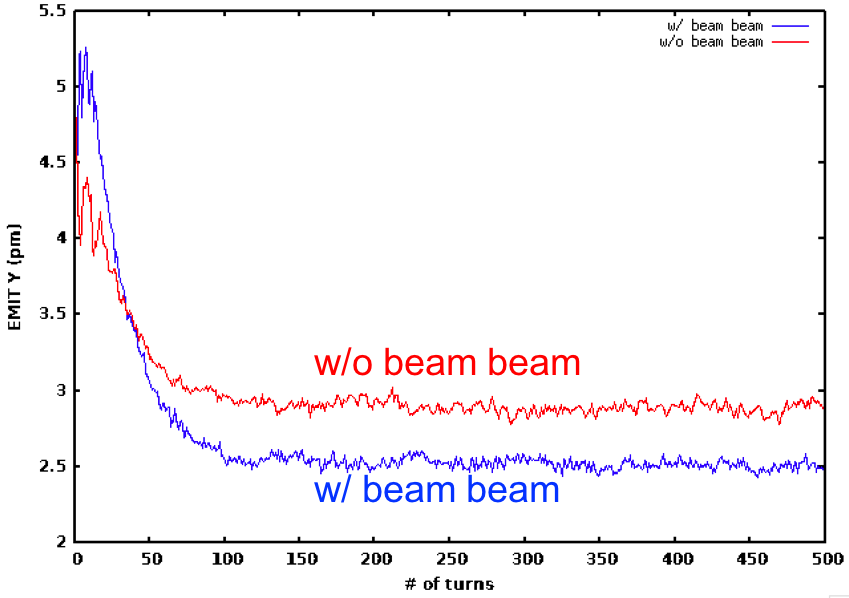}
         \caption{The evolution of the vertical emittance with (blue) and without (red) beam beam element for seed 13  }
  \label{fig:seed13}
    \end{figure}

\begin{figure}[!tbh]
        \includegraphics[width=\columnwidth]{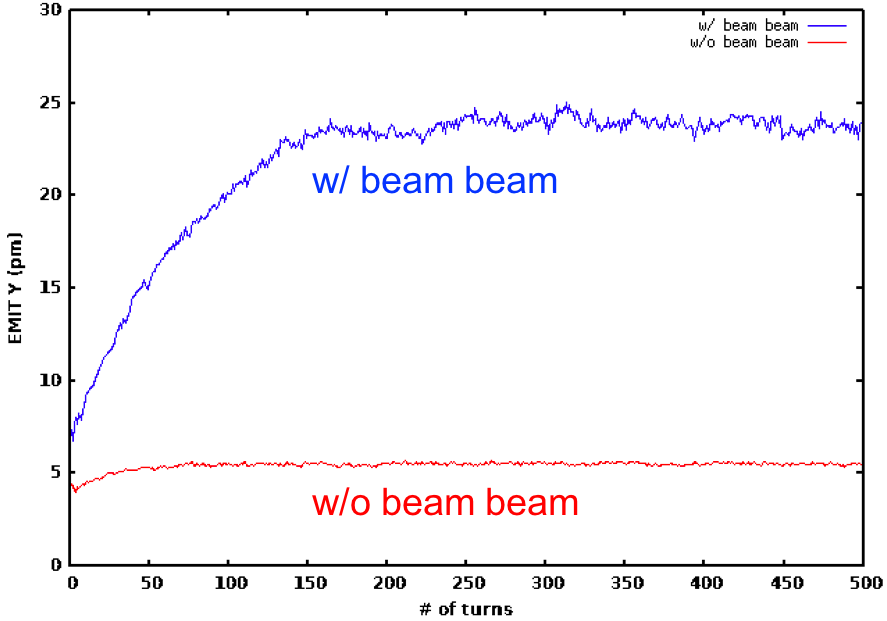}
         \caption{The evolution of the vertical emittance with (blue) and without (red) beam beam element for seed 25  }
  \label{fig:seed25}
    \end{figure}

\begin{figure}[!tbh]
        \includegraphics[width=\columnwidth]{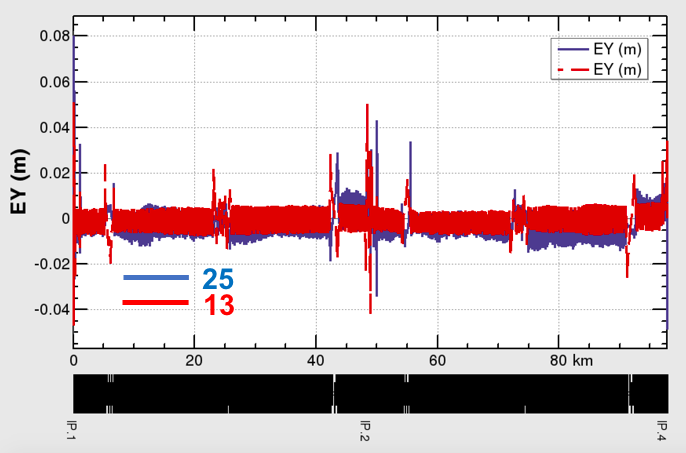}
         \caption{The vertical dispersion in the lattice for seed 13 in red and seed 25 in blue}
  \label{fig:dispseed}
    \end{figure}

\begin{figure}[!tbh]
        \includegraphics[width=\columnwidth]{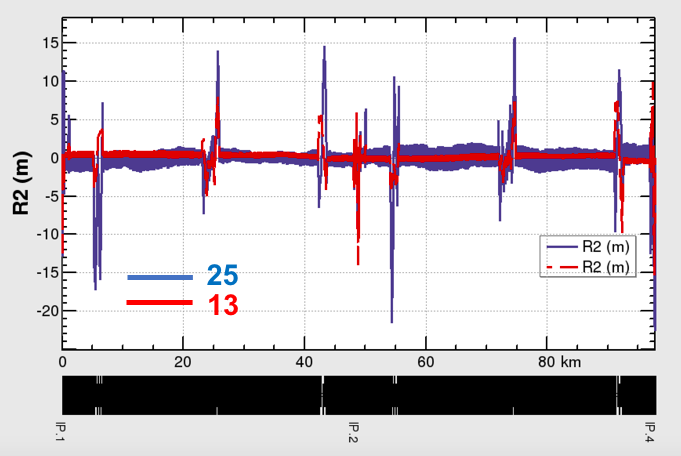}
         \caption{The R2 coupling parameter in the lattice for seed 13 in red and seed 25 in blue  }
  \label{fig:coupseed}
    \end{figure}

Among the twelve different seeds simulated we choose the best and the worst seed and we compare their optics. The best seed is number 13 which didn't result in a vertical emittance blowup in both cases, with and without beam beam (Fig. \ref{fig:seed13}), however the largest blowup was induced by seed number 25 (Fig. \ref{fig:seed25}). We compare the vertical dispersion functions and R2 coupling parameters in the lattice for the two seeds since the vertical emittance depends strongly on these parameters. We notice that the dispersion functions are globally similar for both seeds (Fig. \ref{fig:dispseed}), however the coupling parameter R2 is larger for seed 25 than for seed 13 (Fig. \ref{fig:coupseed}). The vertical dispersions and R2 coupling values are also checked at the IP, and results are reported in Table. \ref{table:dispcoup}. As predicted, R2 values at IPs are higher for seed 25 than for seed 13 which is obviously resulting in a vertical emittance blowup. \newline
Other simulations have been carried on, where the vertical emittance blowup was found to be dependent on the symmetry of the skew quads and the values of the betatron and synchrotron tunes \cite{oide}.
    
    \begin{table}[!tbh]
   \centering
   \begin{tabular}{lcc}
       \toprule
       Seed & $\eta_y$ at IP1 / IP2 ($\mu$m)                      & R2 at IP1 / IP2 (mm)\\
       \midrule
           13        &    -15 / 17     &  0.15 / -0.03       \\ %[3pt]
           25     &        5.3 / -22.1    &    0.5 / 1 \\ %[3pt]
        \bottomrule
   \end{tabular}
   
   \caption{Dispersion and coupling values at IP1 and IP2  for seeds 13 and 25}
   \label{table:dispcoup}
\end{table}

\section{CONCLUSION}
Beam blowup was observed by tracking with and without beam beam in the presence of xy coupling sources in the ring represented by vertical misalignments of sextupoles. This blowup was observed to be dependent of the random generator used for the sextupole misalignments. High R2 coupling parameter values seemed to be responsible for such a blowup. Further simulations are undergoing considering corrected lattice, however this blowup will set further conditions on the low emittance tuning and the choice of the tunes.

\section{ACKNOWLEDGEMENTS}
The authors thank M. Benedikt, A. Blondel, M. Boscolo, E. Levichev, K. Ohmi, D. Shatilov, D. Zhou and the entire FCC-ee Collaboration Team for encouraging the research, useful discussions, and suggestions.

%
% only for "biblatex"
%
\ifboolexpr{bool{jacowbiblatex}}%
	{\printbibliography}%
	{%
	% "biblatex" is not used, go the "manual" way
	
	%\begin{thebibliography}{99}   % Use for  10-99  references
	
} % end \ifboolexpr
%
% for use as JACoW template the inclusion of the ANNEX parts have been commented out
% to generate the complete documentation please remove the "%" of the next two commands
% 
%\newpage

%\include{annexes-A4}

\end{document}